# Properties of $Nb_xTi_{(1-x)}N$ thin films deposited on 300 mm silicon wafers for upscaling superconducting digital circuits.

Author Names: Daniel Pérez Lozano[1], Jean-Philippe Soulié[1], Blake Hodges[2], Xiaoyu Piao[1], Sabine O'Neal[2], Anne-Marie Valente-Feliciano[3], Quentin Herr[1,2], Zsolt Tőkei[1], Min-Soo Kim[1] and Anna Herr[1,2].

[1] imec, Kapel reef 75, B-3001 Heverlee, Belgium
[2] imec USA-Florida, 194 NeoCity Way, Kissimmee, FL 32744, USA
[3] Jefferson Lab, 12000 Jefferson Ave, Newport News, VA 23606, USA
Email:
Daniel.PerezLozano@imec.be

## Abstract

Scaling superconducting digital circuits requires fundamental changes in the current material set and fabrication process. The transition to 300 mm wafers and the implementation of advanced lithography are instrumental in facilitating mature CMOS processes, ensuring uniformity, and optimizing the yield. This study explores the properties of $Nb_xTi_{(1-x)}N$ films fabricated by magnetron DC sputtering on 300 mm Si wafers. As a promising alternative to traditional Nb in device manufacturing, $Nb_xTi_{(1-x)}N$ offers numerous advantages, including enhanced stability and scalability to smaller dimensions, in both processing and design. As a ternary material, $Nb_xTi_{(1-x)}N$ allows engineering material parameters by changing deposition conditions. The engineered properties can be used to modulate device parameters through the stack and mitigate failure modes. We report characterization of $Nb_xTi_{(1-x)}N$ films at less than 2% thickness variability, 2.4% $T_c$ variability and 3% composition variability. The films material properties such as resistivity (140-375 $\Omega \cdot cm$) and critical temperature $T_c$ (4.6 K - 14.1 K) are correlated with stoichiometry and morphology of the films. Our results highlight the significant influence of deposition conditions on crystallographic texture along the films and its correlation with $T_c$.

## Introduction

Superconducting digital logic has the potential to provide high computing density and energy-efficient alternatives [1] for sustainable scaling of exascale and post-exascale high-performance computing (HPC). However, current state-of-the-art superconducting digital circuits suffer from limitations in the feature size, layer count in the fabrication process, power distribution, and mitigation of failure modes [2]. These limitations stem, in part, from the use of Nb as the base material [3]. Despite having the highest critical temperature of all elementary superconductors [4] Nb critical temperature degrades upon annealing [5]. This imposes a thermal budget (150 °C – 200 °C) for integration of multiple metal layers [6] thus, preventing the use of back end of line (BEOL) integration procedures commonly use in CMOS technology (450 °C -500 °C) [7,8]. To overcome these limitations fundamental changes in the material set and processing need to be implemented. $Nb_xTi_{(1-x)}N$ is an alternative material that has been explored for circuit fabrication by the superconducting quantum computing [9–11] and superconducting RF communities [12–14]. This material can reach critical temperatures ($T_c$) up to 17.3 K [15] and it does not degrade [16,17] when subjected to temperatures similar to the ones used for BEOL integration. In contrast to other superconducting nitrides, $Nb_xTi_{(1-x)}N$ exhibits a lower penetration depth, approximately 200-300 nm [18], and higher critical current density $J_c \approx$ 100-140 mA/$\mu m^2$ [19,20]. Furthermore, $Nb_xTi_{(1-x)}N$ showcases consistent superconducting properties even at reduced dimensions as demonstrated by 50×50 $nm^2$ NbTiN superconducting wires sustaining a critical temperature of up to 12.5 K [21]. Moreover, film properties like resistivity, stress and $T_c$ [22,23] can be significantly tuned depending on the deposition conditions This property renders it a promising contender for applications in superconducting digital electronics and the scaling up of these circuits. However, achieving consistency in thickness, resistivity, and critical temperature while being able to engineer these parameters is paramount when it comes to upscaling such circuits.

In this study, we demonstrate the fabrication of $Nb_xTi_{(1-x)}N$ films on 300 mm Si wafers at a deposition temperature of 420 ºC, using different underlying substrates and employing a large variety of sputtering conditions, including $N_2$ flow, Ar flow and deposition power. As a result, we have achieved a wide range of films with critical temperatures ($T_c$) spanning from 4.63 K to 14.1 K, compositional variations measured by the N/(Nb + Ti) ratio, which extends from 0.55 to 0.92, and resistivity values ranging from 140 µΩ·cm to 375 µΩ·cm. We observe that the main predictor of high $T_c$ is the crystallographic orientation. Furthermore, we assess the consistency of the aforementioned parameters across the entire wafer successfully achieving remarkably tight variability, with film thickness varying by 1.3%, $T_c$ by only 2.4%, the N/(Nb + Ti) ratio showing a minimal variability of just 3%. Nevertheless, we have observed variations of resistivity up to 17% which are linked to crystallographic orientation changes across the wafer radius.

# Methods and sample fabrication

$Nb_xTi_{(1-x)}N$ films were deposited on 775µm double side polished (100) Si wafers by physical vapor deposition (PVD). In total seven different samples were prepared, as indicated in Table 1. Silicon oxide (100 nm) was thermally grown on five of the wafers prior to $Nb_xTi_{(1-x)}N$ deposition while in the two remaining samples 100 nm of silicon oxide was deposited by plasma enhanced chemical vapor deposition (PECVD). Additionally, sample 4 was annealed before $Nb_xTi_{(1-x)}N$ deposition for 4 h at 600 ºC in a $N_2$ atmosphere, and both sample 3 and sample 4 were annealed after $Nb_xTi_{(1-x)}N$ deposition for 4 h at 600 ºC in a $N_2$ atmosphere. Nb and Ti were co-sputtered in the presence of a continuous $N_2$ and Ar flow at 420 ºC. The targets (with a purity of 5N) were pre-sputtered for conditioning for 10 min at different powers in the presence of $N_2$ and Ar mixture. In order to obtain a homogenous film thickness, the substrates were rotated during deposition.

| Name | Oxide type | Post-treatment | Nb power (W) | Ti power (W) | Ar flow (sccm) | $N_2$ flow (sccm) | Deposition rate (nm/s) | Thickness (nm) | $\rho_{centre}$ (µΩ*cm) | $T_{c, centre}$ (K) |
|---|---|---|---|---|---|---|---|---|---|---|
| Sample 1 | Thermal oxide | None | 1252 | 750 | 21 | 18 | 0.06 | 53.6 | 229.8 | 13.1 |
| Sample 2 | Thermal oxide | None | 754 | 500 | 21 | 15 | 0.03 | 46.0 | 291.2 | 12.6 |
| Sample 3 | PECVD oxide | 600 ºC | 754 | 500 | 21 | 15 | 0.03 | 48.3 | 385.8 | 11.5 |
| Sample 4 | PECVD oxide + 600 ºC | 600ºC | 754 | 500 | 21 | 15 | 0.03 | 48.0 | 351.7 | 11.4 |
| Sample 5 | Thermal oxide | None | 1169 | 700 | 21 | 15 | 0.13 | 52.9 | 185.4 | 9.7 |
| Sample 6 | Thermal oxide | None | 1252 | 750 | 21 | 16 | 0.12 | 47.0 | 184.4 | 5.9 |
| Sample 7 | Thermal oxide | None | 1252 | 750 | 21 | 15 | 0.16 | 49.1 | 186.1 | 4.6 |

*Table 1. Deposition conditions and film property values.*

Crystal structure of the films and thickness were determined respectively by X-Ray Diffraction (XRD) and X-Ray reflectance (XRR) using Cu-Kα radiation in Panalytical X'Pert diffractometer. The resistivity (ρ) of the films was determined by measuring the sheet resistance on full wafers with a KLA Tencor RS100 four-point prober in combination with the XRR film thickness. Composition of the films was determined by using a combination of Rutherford backscattering spectrometry (RBS) and Elastic Recoil Detection (ERD) measurements. RBS used a 6SDH Pelletron accelerator (National Electrostatics Corporation, NEC) where the He+ ion beam is tuned at 1.523 MeV. ERD measurements used a 6SDH Pelletron accelerator (National Electrostatics Corporation, NEC) and a primary ion beam of $Br^{4+}$ accelerated to 9.6 MeV. The 4-point probe method was used to measure $T_c$ of the films. Samples were first loaded and cooled down in a cryostat to 2.6 K. The temperature was then slowly increased to

15 K using a heater. While the temperature was increasing, a source-measure unit performed 4-point measurements on the samples. A probing current was passed across two pins contacting the sample, and two neighbouring pins were used to measure the voltage generated. This probing measurement was performed ten times for each temperature that was measured. The resulting voltages were averaged and plotted against temperature to extract the critical temperature.

## Results and discussion

Room temperature resistivity and critical temperature were measured at the centre of each film (Figure 1). Notably, no discernible correlation between these two parameters was evident. It was observed that low resistivity values correspond to low critical temperatures, followed by a subsequent increase in $T_c$ along with resistivity. Interestingly, for the highest measured $T_c$, resistivity exhibited a subsequent decrease.

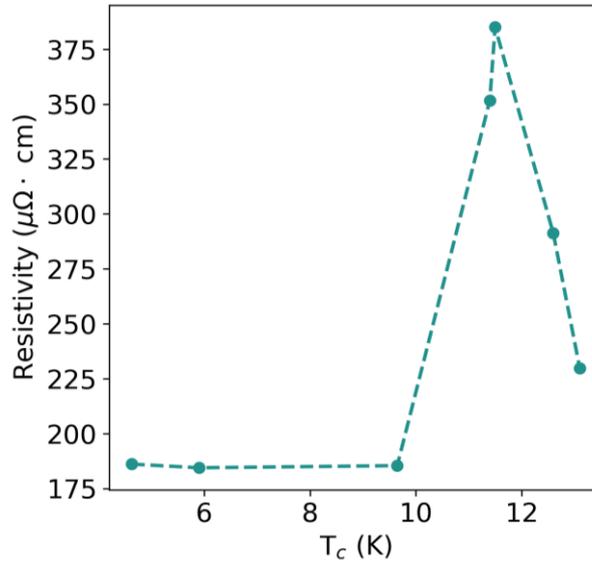

Figure 1. Room temperature resistivity as a function of $T_c$.

Next, we examine the crystal structure at the same locations where room temperature resistivity and $T_c$ were measured (Figure 2(a)). Overall, there is no significant differences in the crystal structure of the different films. All of them show three peaks corresponding with (111), (200), and (220) orientations, being all films textured along the (111) direction. Notably, our observations reveal that films with lesser texturing along the (111) direction tend to exhibit higher $T_c$ ( Figure 2(b)-(c)). A similar trend has been observed by Ge et *al*. after annealing $Nb_xTi_{(1-x)}N$ films at different temperatures [16] and by Savvides et *al*. when depositing TiN films at temperatures higher than 500 ºC [24].

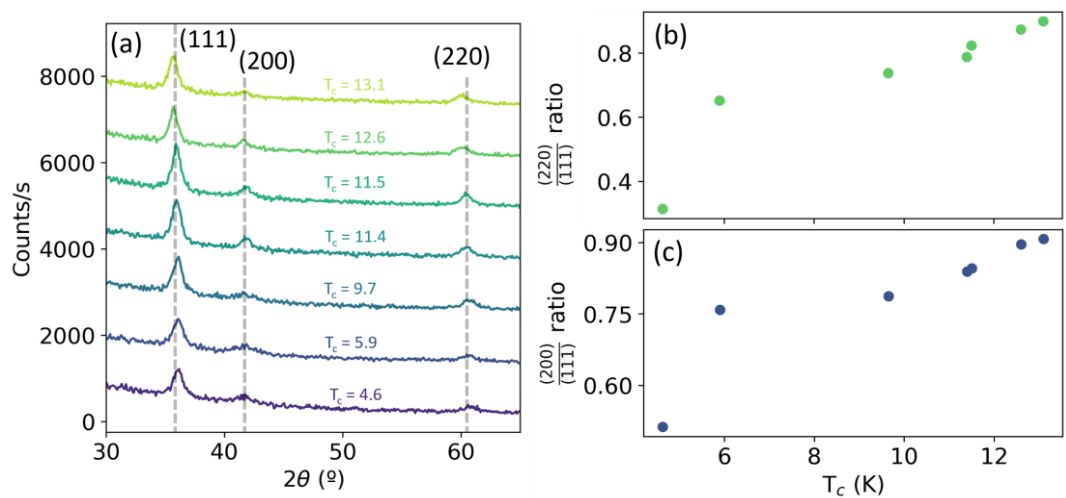

Figure 2. (a) XRD spectrum at the centre of each film ordered by increasing critical temperature. (b) (220):(111) peak height ratio as a function of critical temperature. (c) (200):(111) peak height ratio as a function of critical temperature. The data on the plots is ordered according to order of samples in Table 1.

Note that sample 2, 3 and 4 have different $T_c$ despite using exactly the same deposition parameters. This points out the effect of the underlying substrate on $T_c$. Even when using the same underlying material substrate, the deposition technique and processing affects the crystal structure of the $Nb_xTi_{(1-x)}N$ film deposited on top.

Finally, we examine the proportions of Nb, Ti and N present at the centre of each film (Figure 3(a-c)). In general, films with less atomic concentration of Nb and Ti, and higher atomic concentration of N exhibit higher $T_c$. Previous studies by Maezawa et al. [25] have shown that N/(Nb+Ti) ratios higher than one result in $Nb_xTi_{(1-x)}N$ films with higher $T_c$, which is in agreement with our observations (Figure 3(d)). Despite the different deposition conditions and substrate types, the only parameters that are strongly correlated with critical temperature are the XRD peak ratios and the atomic concentrations of Nb, Ti and N. Thus, to disentangle how these parameters interplay, we study their distribution across the entire wafer for the two films with the highest $T_c$ measured at the centre: sample 1 and sample 2.

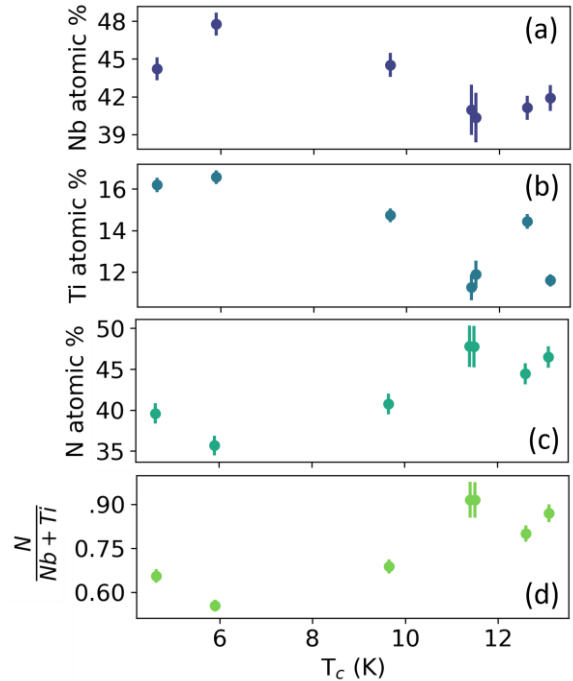

Figure 3. (a-c) Nb, Ti and N atomic concentrations for each of the films fabricated, together with N/(Nb + Ti) ratio (d). The data on the plots is ordered according to order of samples in Table 1.

The resistivity maps show a concentric pattern with the highest resistivity values at the centre and the lowest at the edges (Figure 4), although the variability ($C_v = 100 * \frac{\sigma_\rho}{\langle\rho\rangle}$) changes significantly from wafer to wafer, ranging between 2.4% in sample 3 to 14.3% in sample 1 and 17.7% in sample 2. This suggests that the concentric pattern is caused by the plasma non-uniformity rather other elements like thermal gradients in the wafer holder. A similar pattern has also been reported in $Nb_xTi_{(1-x)}N$ films deposited on 100 mm sapphire wafers [26], although the effect was explained because the deposition system used was originally intended for 50 mm wafers. Variations on film thickness across the wafer are smaller than 2%, therefore they cannot explain the large variations observed in resistivity. It has also been previously shown that resistivity is correlated with $T_c$ in $Nb_xTi_{(1-x)}N$ films [27], as well as in NbN [28] films and TiN [29] films, and that lower resistivity films results in higher $T_c$. Thus, we expect the edges of our samples to have $T_c$ than the centre.

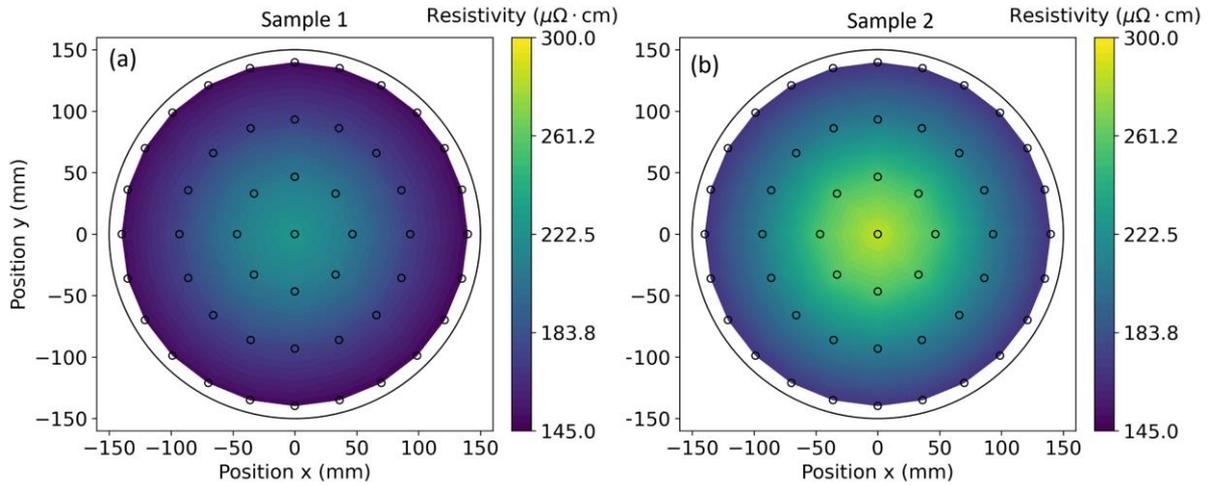

Figure 4. Resistivity map for sample 1(a) and sample 2(b).

Next, we examine the crystal structure at three different points across the radius of the wafer: r = 0 mm, 70 nm and 130 mm (Figure 5(a-b)). Both samples are textured along the (111) direction at the centre of the wafer. The intensity of this peak changes along the radius, indicating a shift in the preferred crystal orientation of the film. Drawing from the observed trend in the previously measured films, this also implies that the $T_c$ values at the edges of the wafer are likely to be higher in comparison to the centre.

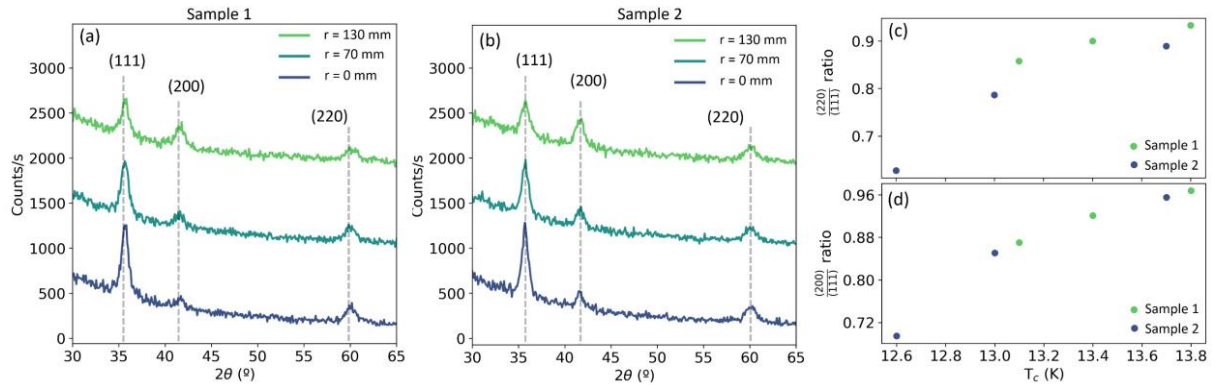

Figure 5. (a-b) XRD spectra at r = 0 mm, r = 70 mm and r = 130 mm for sample 1 and sample 2. (c-d) (200):(111) and (220):(111) XRD peak ratios as a function of $T_c$ measured at the same locations.

Critical temperature of the films was measured at 9 different locations (Figure 6). Sample 1 shows $T_c$ ranging from 13.1 – 14.1 K with a coefficient of variation of 2.4 %. $T_c$ in sample 7 ranges between 12.6 K - 13.7 K, with coefficient of variation of 2.8 %. In both films the area with the higher $T_c$ happens at the edge of the wafer, as we hypothesized, corresponding with the lowest room temperature resistivity and with the highest (200):(111) and (220):(111) XRD peak ratios as shown in Figure 5(c-d). This observation is consistent with previous reports on the XRD spectrum and $T_c$ change of $Nb_xTi_{(1-x)}N$ films [16,17].

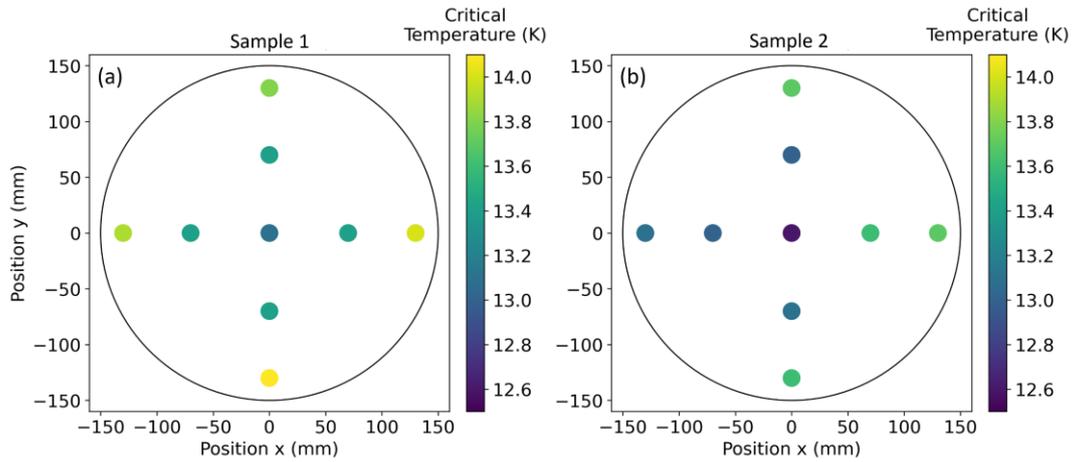

Figure 6. $T_c$ wafer map of sample 1 (a) and sample 2 (b).

Finally, Nb, Ti and N atomic concentrations (Figure 7) were determined in the same areas where $T_c$ was measured. Sample 1 contains slightly more Nb and N than sample 2. The Nb, Ti and N atomic concentrations variability in sample 1 are 1.7%, 1.9% and 0.9% respectively. In sample 2 the Nb, Ti and N atomic concentrations variability are 2.1%, 2.6% and 1.7%. Overall, there is no correlation between $T_c$ and stoichiometry of the film in the different locations measured (Figure 8). This suggests that within the range of values measured the atomic concentration of Nb, Ti and N do not affect $T_c$ of the films.

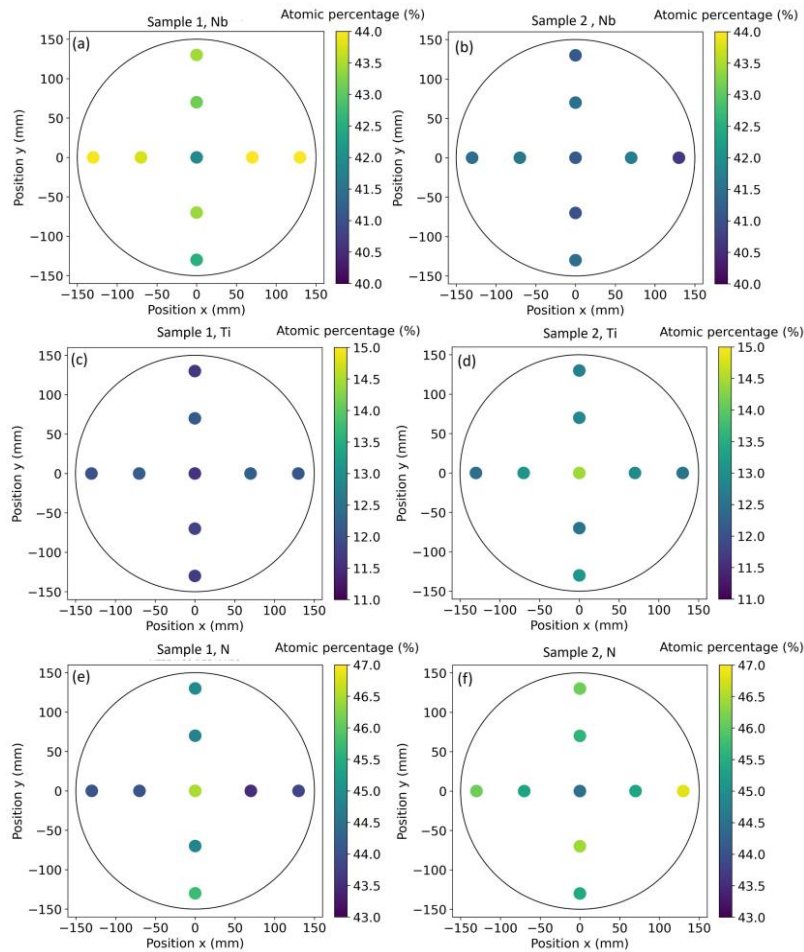

Figure 7. (a,c,e) atomic concentration wafer map for Nb, Ti an N of Sample 1. (b, d, f) atomic concentration wafer map for Nb, Ti an N of Sample 2.

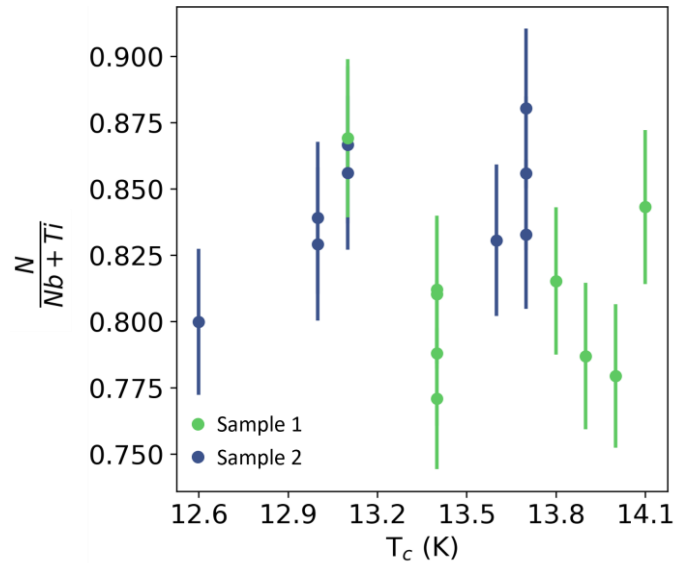

Figure 8. N/(Nb + Ti) ratio as a function of $T_c$ at different wafer locations for sample 1 and sample 2.

It is important to note that there are some apparent discrepancies between the observed trends in the central points of samples 1-7 and the trends noted across the radial profile of sample 1 and sample 2, particularly concerning the relationship between $T_c$ and resistivity. While there seems to be no correlation between resistivity and $T_c$ when examining the central points of samples 1-7 (Figure 1), a correlation emerges when considering values across the radial profile of the wafer. There are significant differences between Nb, Ti and N atomic concentrations of the central points of the different films, but there is no such a difference between the atomic concentration of the different points across the radius for the same wafer. These compositional distinctions impact the lattice constant of the crystals [30], subsequently influencing grain size, grain boundaries and resistivity. Consequently, it becomes evident that the most relevant parameters for predicting high $T_c$ in our films are the (200):(111) and (220):(111) XRD peak ratios.

## Conclusion

We have deposited $Nb_xTi_{(1-x)}N$ films at 420 ºC on 300 mm Si wafers covering a broad spectrum of $T_c$, resistivity, and composition values. By conducting a comparative analysis of these properties both across wafers and within each individual wafer, we have established a correlation between $T_c$ and a singular parameter: the (200):(111) and (220):(111) XRD peak ratios. This correlation is independent of deposition conditions and substrate type and reveals that higher peak ratios correspond to higher $T_c$ values. Furthermore, our understanding of the interplay between resistivity, composition, and deposition parameters offers the prospect of tailoring $Nb_xTi_{(1-x)}N$ films to meet specific requirements. Lastly, we have achieved thickness variability values below 2% and $T_c$ variability values below 2.4% on 300 mm wafers. This development paves the way for the integration and upscaling of superconducting digital electronic devices using alternative materials.

## Acknowledgements

A-M Valente-Feliciano is supported by the U.S. Department of Energy, Office of Science, Office of Nuclear Physics under contract DE-AC05-06OR23177. Work at imec and imec USA is supported by imec INVEST+ and by Osceola County. The authors would like to thank Johan Desmet and Johan Meersschaut for the RBS and ERD measurements.